\documentclass[prl,aps,twocolumn,superscriptaddress,showpacs]{revtex4}
\usepackage{graphicx}
\usepackage{amsmath}

\begin{document}

\title{Enhanced resolution of lossy interferometry by coherent amplification of single photons}
\author{Chiara Vitelli}
\affiliation{Dipartimento di Fisica, Sapienza Universit\`{a} di Roma,
piazzale Aldo Moro 5, I-00185 Roma, Italy}
\affiliation{Consorzio Nazionale Interuniversitario per le Scienze Fisiche
della Materia, piazzale Aldo Moro 5, I-00185 Roma, Italy}

\author{Nicol\`{o} Spagnolo}
\affiliation{Dipartimento di Fisica, Sapienza Universit\`{a} di Roma,
piazzale Aldo Moro 5, I-00185 Roma, Italy}
\affiliation{Consorzio Nazionale Interuniversitario per le Scienze Fisiche
della Materia, piazzale Aldo Moro 5, I-00185 Roma, Italy}

\author{Lorenzo Toffoli}
\affiliation{Dipartimento di Fisica, Sapienza Universit\`{a} di Roma,
piazzale Aldo Moro 5, I-00185 Roma, Italy}

\author{Fabio Sciarrino}
\email{fabio.sciarrino@uniroma1.it}
\affiliation{Dipartimento di Fisica, Sapienza Universit\`{a} di Roma,
piazzale Aldo Moro 5, I-00185 Roma, Italy}
\affiliation{Istituto Nazionale
di Ottica, largo Fermi 6, I-50125 Firenze, Italy}

\author{Francesco De Martini}
\affiliation{Dipartimento di Fisica, Sapienza Universit\`{a} di Roma,
piazzale Aldo Moro 5, I-00185 Roma, Italy}
\affiliation{Accademia Nazionale dei Lincei, via della Lungara 10, I-00165 Roma,
Italy}

\begin{abstract}
In the quantum sensing context most of the efforts to design
novel quantum techniques of sensing have been constrained to idealized,
noise-free scenarios, in which effects of environmental
disturbances could be neglected. In this work, we propose
to exploit optical parametric amplification to boost
interferometry sensitivity in the presence of losses in a minimally invasive scenario. By
performing the amplification process on the microscopic probe
after the interaction with the sample, we can beat the losses
detrimental effect on the phase measurement which affects the
single-photon state after its interaction with the sample, and thus improve the
achievable sensitivity.
\end{abstract}


\maketitle

The aim of quantum sensing is to develop methods
to extract the maximum amount of information from a system with a
minimal disturbance on it. Indeed, the possibility of performing
precision measurements by adopting quantum resources can increase
the achievable precision going beyond the semiclassical regime of
operation \cite{Giov04b,Giov06,Hels76}. 
%The employ of a quantum probe and 
%entangled measurement schemes in order to
%estimate a classical parameter can beat the standard quantum limit imposed
%on the accuracy of the measurement \cite{Hels76}. 
In the case of
interferometry, this can be achieved by the use of the so-called N00N states,
which are quantum mechanical superpositions of just two terms, corresponding
to all the available photons $N$ placed in either the signal arm or the
reference arm. The use of N00N states can enhance the precision in phase estimation to $1/N$, thus
improving the scaling of the achievable precision with respect to the employed
resources \cite{Boto00,Dowl08}. This approach can have wide applications for
minimally invasive sensing methods in order to extract the maximum amount of
information from a system with minimal disturbance. 
%Imaging of biological samples 
%and of an ancient artifact are examples of situations where it is clearly beneficial 
%to use as weak light as necessary to achieve a desired level of measurement precision.
%In the quantum domain there is an even stronger motivation for minimally invasive 
%measurements since the back action of the measurement actually changes the state of 
%the quantum system under investigation. 
The experimental
realization of protocols involving N00N states containing up to 4 photons
have been realized in the last few years \cite{Dang01,Walt04,Mitc04,Eise05,Naga07a}. 
Other approaches \cite{Ono10,Afek10} have focused on exploiting coherent and squeezed 
light to generated fields which approximate the features of N00N states. Nevertheless, 
these quantum states turn out to be extremely fragile under losses and
decoherence \cite{Gilb08}, unavoidable in experimental implementations.
A sample, whose phase shift is to be measured, may at the same time
introduce high attenuation. Since quantum-enhanced modes of operations
exploit fragile quantum mechanical features the
impact of environmental effects can be much more deleterious than in
semiclassical schemes, destroying completely quantum benefits \cite{Rubi07,Shaj07}.
This scenario puts the beating of
realistic, noisy environments as the main challenge in developing quantum
sensing. Very recently, the theoretical and experimental investigations of
quantum states of light have attracted much attention, leading to the best
possible precision in optical two-mode interferometry, even in the presence of
experimental imperfections \cite{Huve08,Dorn09,Macc09,Demk09,Kacp09,Lee09}.\\
\begin{figure}[t!!]
\includegraphics[width=0.49\textwidth]{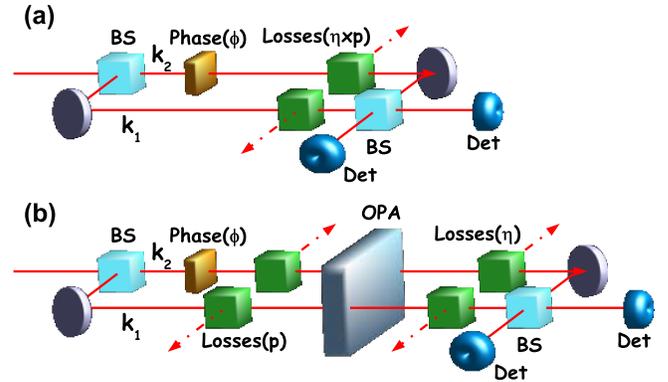}
\caption{Scheme for the phase measurement. (a) Interferometric scheme
adopted to estimate the phase ($\phi$) introduced in the mode $k_{2}$.
(b) Interferometric scheme adopting a single photon and the optical
parametric amplifier: the amplification of the single photon state is performed
before dominant losses.}
\label{fig:schema_concettuale}
\end{figure}
In this work, we adopt a hybrid approach based on a high
gain optical parametric amplifier operating for any polarization
state in order to transfer quantum properties of different
microscopic quantum states in the macroscopic regime
\cite{DeMa98,DeMa98a}. By performing the amplification process of the
microscopic probe after the interaction with the sample we can
beat the losses detrimental effect on the phase measurement which
affects the single-photon state after the sample. Our approach
can be adopted in a minimally invasive scenario where a fragile sample,
such as biological or artifacts systems, requires as few photons as 
possible impinging on it in order to prevent damages. The action of
the amplifier, i.e., the process of \emph{optimal phase covariant
quantum cloning}, is to broadcast the phase information codified
in a single photon into a large number of particles. Such
multiphoton states have been shown to exhibit a high resilience to
losses \cite{DeMa08,DeMa09,DeMa09a} and can be manipulated by exploiting a
detection scheme which combines features of discrete and
continuous variables. The effect of losses on the macroscopic
field consists in the reduction of the detected signal and not in
the complete cancellation of the phase information as would happen
in the single-photon probe case, thus improving the achievable
sensitivity. This improvement does not consist in a scaling factor
but turns out to be a constant factor in the sensitivity depending
on the optical amplifier gain. Hence, the sensitivity still scales 
as the $\sqrt{N}$, where $N$ is the number of photons impinging 
on the sample, but the effect of the amplification process is 
to reduce the detrimental effect of losses by a factor proportional 
to the number of generated photons.\\
\indent Let us review the adoption of single photons in order
to evaluate the unknown phase $\varphi $, Fig.1-(a). The phase
$\varphi $ introduced in the path $k_{2}$ is probed by sending to
the sample $N$ input photons, each one in the state
$2^{-1/2}\left( \left| 1\right\rangle _{k1}+\left| 1\right\rangle
_{k2}\right) $. After the propagation, the sample introduces a
phase $\varphi $ on the probe beam and each photon is found in the
state: $\frac{1}{\sqrt{2}}\left( \left| 1\right\rangle
_{k1}+e^{i\varphi }\left| 1\right\rangle _{k2}\right) $. The two
modes $k_{1} $ and $k_{2}$ are then combined on a beamsplitter
($BS$) and detected by ($D'_{1},D'_{2}$) with an overall detection
efficiency equal to $t$. $N$ performed experiments leads to an
output signal equal to $I= I(D'_{1}) - I(D'_{2}) = t
N\cos\varphi$, whose fluctuations are given by $\sigma =\left( t N
\right) ^{1/2}$. The uncertainty on the phase measurement around
the value $\frac{\pi }{2}$ can hence be estimated as $\Delta
\varphi =\left( \frac{\partial I}{\partial \varphi }\right)
^{-1}\triangle I =\frac{1}{ \sqrt{ t N}}$, the semiclassical shot noise
limit, and the sensitivity of the interferometer can be evaluated
as $S_{\mathrm{1phot.}}=\frac{1}{\Delta \varphi} = \sqrt{t N}$.\\
\indent In order to avoid the detrimental effect of a low value of $t$, our
strategy involves the amplification of the single-photon probe, Fig.1-(b). In
the theory and experiment described here, the two modes $k_{1}$
and $k_{2}$ correspond to two orthogonal polarization modes:
horizontal ($H$) and vertical ($V$) associated to the same
longitudinal spatial mode $k$. The input
single photon is prepared in the polarization state: $|+\rangle =\frac{1}{%
\sqrt{2}}\left( |H\rangle +|V\rangle \right) $. After the propagation over
the interferometer, the photon acquires the unknown phase $\varphi $: $%
|\varphi \rangle =\frac{1}{\sqrt{2}}\left( |H\rangle +e^{\imath \varphi
}|V\rangle \right) $. The amplification performed by the optical parametric device
generates the output state $\label{eq:output_int_OF_2}|\Phi ^{\varphi
}\rangle =\widehat{U}_{OPA}|\varphi \rangle =\cos \frac{\varphi }{2}|\Phi
^{+}\rangle +\imath \sin \frac{\varphi }{2}|\Phi ^{-}\rangle $, where $|\Phi
^{+,-}\rangle $ are the wavefunctions described in Ref. \cite{DeMa08}.
Precisely, the state $|\Phi ^{+}\rangle $ ($|\Phi ^{-}\rangle $) presents a
Planckian probability distribution as a function of photons polarized $\vec{\pi}%
_{-}$ ($\vec{\pi}_{+}$) and a long tail distribution as a function of
photons polarized $\vec{\pi}_{+}$ ($\vec{\pi}_{-}$). The two distributions
belonging to the state $|\Phi ^{+}\rangle $ and $|\Phi ^{-}\rangle $
partially overlap, but become distinct on the border of the Fock states
plane \cite{Naga07}. For the state $|\Phi ^{\varphi }\rangle ,$ the average
number of photons emitted over the polarization mode $\vec{\pi}_{+}$ is
equal to $\langle n_{+} \rangle =\overline{n}+\cos ^{2}\frac{\varphi }{2}(2\overline{n}+1)$ with $%
\overline{n}=$sinh$^{2}g$ and $g$ the gain of the amplifier, while the
average number of photons emitted over the polarization mode $\vec{\pi}_{-}$
is equal to $\langle n_{-} \rangle =\overline{n}+\sin ^{2}\frac{\varphi }{2}(2\overline{n}+1)$.
The previous expressions lead to a phase-dependent intensity with a visibility
$V=\frac{\left<n_{+}\right>-\left<n_{-}\right>}{\left<n_{+}\right> +\left<n_{-}\right>} 
\rightarrow 0.50$ for $g\rightarrow\infty$. The resilience to losses of such
multiphoton fields \cite{DeMa09} renders them suitable for the implementation of
quantum information applications in which noisy channels and low detection
efficiency are involved. We consider the case in which the losses are
unavoidable during the detection process, and happen after the single-photon
amplification (Fig.\ref{fig:schema_concettuale}). After the propagation over
a lossy channel, the state evolves from $\label{eq:evolution_state_losses}%
|\Phi ^{\varphi }\rangle \,\langle \Phi ^{\varphi }|$ into a mixed state $\hat{\rho}%
_{\eta }^{\varphi }$. For details on the explicit expressions of
the coefficients of the density matrix $\hat{\rho}_{\eta
}^{\varphi }$, see \cite{DeMa09}.

After the amplification stage and the transmission losses, the
received field is analyzed through single-photon detectors ($D'_{1},D'_{2}$) in the $%
\left\{ \vec{\pi}_{+},\vec{\pi}_{-}\right\} $ polarization basis.
Our aim is to compare the achievable sensitivity \textsl{with} and
\textsl{without} the optical amplifier ($g=0$). To take into account
experimental imperfections, we divide the losses $t$ in two
contributions: the first one includes all the losses between the
sample and the optical amplifier ($p$), while the
second parameter
takes into account all the inefficiencies up to the detection stage ($\eta $):
$t=p\times\eta$. Our strategy cannot compensate for losses that occur 
before the amplifier ($p$), but can compensate for large (even very large, 
if the gain is high enough) losses after the amplification ($\eta$).
A first insight on this property of the optical parametric amplifier
has been given in Ref.\cite{Leve97} by analyzing the signal-to-noise
of the amplification of a coherent state signal in lossy conditions.
The sensitivity $S_{ampl.}$, obtained by measuring the
difference $\langle D \rangle = \langle n_{+} \rangle - \langle n_{-} \rangle$
intensity signals provided by the detectors around the phase value
$\varphi=\frac{\pi}{2}$, is found to be:
\begin{equation}
\label{eq:sensitivity_OPA_single_photon}
S_{ampl.} =  \frac{\sqrt{N}p \eta c}{\left\{ \eta^{2}
\left[p\overline{n}(4c + 2)
+ 2 \overline{n} c \right] + \eta [p c +
2 \overline{n} ]\right\}^{1/2}}
\end{equation}
with $c=2 \overline{n} + 1$.\\
\begin{figure}[ht!]
\includegraphics[width=0.49\textwidth]{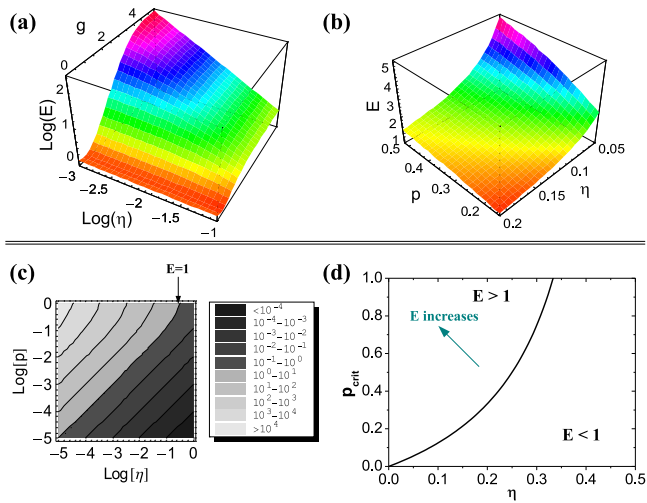}
\caption{\textbf{(a)} Logarithm of the enhancement versus the
nonlinear gain $g$ and the transmittivity of the lossy channel
$\eta$ for a value of losses between the phase-shifter 
and the amplifier equal to $p=0.5$. \textbf{ (b)}  Trend of $E$ as a 
function of losses before amplification $0.15 \leq p \leq 0.5$ and after amplification 
$0.05\leq \eta \leq 0.2$ ($g=4.5$). \textbf{(c)} Contour plot of the enhancement as a 
function of the logarithm of $p$ and $\eta$. The lighter region corresponds 
to $E>1$, the darker one to $E<1$. \textbf{(d)} Non ideal case $p \neq 1$ : 
trend of the injection probability critical value for which $E > 1$ as a function 
of the detection efficiency.} 
\label{fig:enhanced}
\end{figure}
Let us first consider the case $p=0.5$: Fig.2-(a) reports the logarithm of the enhancement 
of the squared sensitivity $E=\left(\frac{S_{\mathrm{ampl.}}}{S_{\mathrm{1phot.}}}\right)^{2}$
versus $g$ and $\eta$. $E$ represents the reduction factor in the number of photons sent 
onto the sample in order to obtain the same information on the phase $\varphi$, by exploiting 
the amplification strategy with respect to the single-photon probe scheme. 
As it can be observed in Fig.2-(a), a large improvement can be obtained in the
regime of high losses and large gain of the amplifier. The motivation of such
behavior is the following: the present approach allows us to increase
the number of detected photons by a factor $4\overline{n}$ with respect to
the single-photon case keeping a visibility of the fringe patterns reduced only 
to $50\%$ (for $g\rightarrow \infty$). The enhancement E is then slightly affected by
the reduction in the visibility, due to the amplification noise \cite{Glau66}, while it is 
significantly improved by the increase in the detected signal.\\
\indent In Fig.2-(b) we report the trend of the 
enhancement as a function of losses $0.15 \leq p \leq 0.5$ and $0.05 \leq \eta \leq 0.2$
for a nonlinear gain of $g=4.5$. 
Such a range corresponds to typical values of losses and detection efficiency achievable
in practical schemes. We observe that also in this regime an enhancement 
greater that $1$ can be obtained. For large values of the gain $g$ the enhancement 
saturates to the value: $E_{lim}=\frac{p}{\eta (2p+1)}$ [contour plot in Fig.2-(c)]; we 
can then identify a critical value of $p$ above which the enhancement is greater than $1$: 
$p_{crit}= \frac{\eta}{1-2\eta}$. For $\eta\geq 0.33$ no enhancement can be achieved by 
exploiting the amplification strategy (see figure 2-(d)). \\
\indent The previous theoretical predictions have been experimentally
tested by adopting a high gain optical parametric amplifier with a
maximum gain $g=4.5$. A detailed description of the apparatus can
be found in Ref.\cite{Naga07,DeMa08}. The single-photon probe
was generated in a first nonlinear crystal and heralded by a
trigger detector $D_{T}$; hence, the phase shifting $\varphi$ was
introduced via a Soleil-Babinet compensator. Then the probe photon
was superposed to an ultraviolet pump beam and injected into a
second nonlinear crystal realizing the optical parametric
amplifier. The output radiation was then coupled with a single
mode fiber and detected with single-photon detectors. Additional
controlled losses were introduced by adopting neutral optical
filters. For the sake of simplicity we propose working in the
single-photon counting regime; in order to describe the detection
apparatus in a linear regime, the following condition for the
average number of detected photons must be satisfied $\eta \langle
n_{\pm }\rangle <<1$ (experimental details can be found in the supplementary 
material \cite{supplementary}). We found experimentally a value of $p$
equal to $0.15$ due to spatial and spectral mismatch between the
injected single photon and the ultraviolet pump beam ($k'_{p}$).
The output fringe patterns have been recorded for different values
of the gain $g$ and hence of the generated number of photons in the
amplifier.  In the extreme condition with $\eta=3 \times 10^{-4}$ and
$g=4.5$ we observed an enhancement of a factor $\sim 210$, as
shown in Fig.\ref{fig:experiment}-(a), in which is reported the
trend of $E$ as a function of the amplifier gain, compared with
the theoretical prediction.\\
\begin{figure}[b!!]
\includegraphics[width=0.49\textwidth]{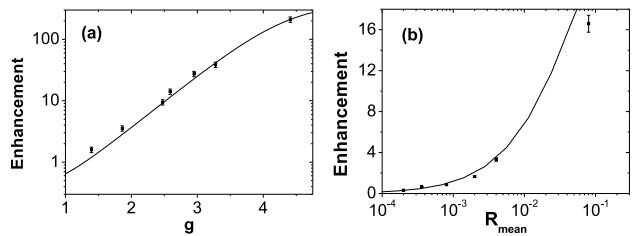}
\caption{\textbf{(a)} Experimental results of the enhancement $E$ versus the
nonlinear gain for the counting detection strategy. Continuous line: 
theoretical prediction for the expected enhancement with $\eta=3 \times 
10^{-4}, p=0.15$. \textbf{(b)} Experimental results of the enhancement $E$ versus the
signal rate, the continuous line reports  the theoretical prediction for $p=0.14$, 
$\eta=0.005$.}
\label{fig:experiment}
\end{figure}
\indent We now discuss the optimality of the measurements performed on the
multiphoton state, in order to extract the maximum information
about the phase $\varphi$ codified in the optical field. This
quantity is expressed by the quantum Fisher information
\cite{Hels76,Pari09}, defined as $H(\varphi) = \mathrm{Tr}[
\hat{\rho}^{\varphi} \hat{L}_{\varphi}]$, where
$\hat{L}_{\varphi}$ is the symmetric logarithmic derivative
$\partial_{\varphi} \hat{\rho}^{\varphi} = \frac{\hat{L}_{\varphi}
\hat{\rho}^{\varphi} + \hat{\rho}^{\varphi} \hat{L}_{\varphi}}{2}$
and $\hat{\rho}^{\varphi}$ is the density matrix of the state in
which the phase is codified. The quantum Cramer-Rao bound
\cite{Hels76} quantifies the maximum precision achievable on the
estimation of the phase $\varphi$ optimized over all possible
measurements as: $\Delta^{2} \varphi \geq 1/H(\varphi)$. In the high lossy regime $\eta
\langle n_{\pm} \rangle \ll 1$, the single-photon amplified states lead to
a quantum Fisher information equal to $H_{ampl}(\varphi) \approx
2\overline{n}\eta p (1+p^{-1})^{-1} $, to be compared with the single-photon case,
which gives $H_{1phot}(\varphi) = \eta p$. This result allows us to
investigate the optimality of the counting measurement strategy.
The sensitivity achieved with this scheme, given by
Eq.(\ref{eq:sensitivity_OPA_single_photon}), can be written in the
high lossy regime as $S^{2}_{ampl} = (\Delta^{2} \varphi)^{-1}
\approx 2\overline{n}\eta p (1+p^{-1})^{-1}$, thus saturating the Cramer-Rao bound
and ensuring the optimality of this scheme
in the high lossy regime. \\
\indent As a more sophisticated strategy, it is possible to elaborate on an
approach which leads to higher visibility of the detected fringe
patterns at the cost of a reduced detection rate of the signal:
the output radiation is measured in polarization with two linear
detectors, for instance photomultipliers. The intensity signals
generated by the detectors proportional to the orthogonally
polarized number of photons are compared shot by shot by the
orthogonality-filter (OF) electronic device introduced in Ref.
\cite{DeMa08}.
When the number of photons $m_{\varphi }$, detected in the $\vec{\pi}%
_{\varphi }$ polarization, exceeds $n_{\varphi _{\bot }}$, detected in the $%
\vec{\pi}_{\varphi \perp }$ polarization, over a certain adjustable
threshold $k$, i.e. $m_{\varphi }-n_{\varphi _{\bot }}>k$, the (+1) outcome
is assigned to the event and the state $|\Phi ^{\varphi }\rangle $ is
detected. On the contrary, when the condition $n_{\varphi _{\bot
}}-m_{\varphi }>k$ is satisfied, the (-1) outcome is assigned and the state $%
|\Phi ^{\varphi \bot }\rangle $ is detected. Finally, an inconclusive result
(0) is obtained when the unbalance between detected pulses does not 
exceed the threshold $k$. As the gain is increased,
the number of transmitted photons $\eta \langle n
\rangle$ becomes sufficient to detect all the $N$ repeated trials. In the high
losses regime, at variance with the single-photon case, all pulses can be
exploited to extract information about the phase $\varphi$. The action of
the OF is then to select those events that can be discriminated with higher
fidelity, leading to an increase in the visibility, at the cost of
discarding part of the data.
%\begin{figure}[h]
%\includegraphics[width=0.4\textwidth]{fig4_enhancement.eps}
%\caption{Experimental results of the enhancement $E$ versus the
%signal rate, the continuous line reports  the theoretical
%prediction for $p=0.14$, $\eta=0.005$.} \label{fig:experiment_OF}
%\end{figure}
According to these considerations, the ``detection'' efficiency of
the scheme, i.e. the percentage of detected events, is given by
the average signal $\overline{\eta} = R_{mean}(k)$ filtered by the
OF device. This parameter $\overline{\eta}$ corresponds to the
overall efficiency of the amplification-OF-based detection scheme.
We calculated the phase measurement uncertainty through the standard definition $%
\Delta \varphi _{OF}=\Delta R_{OF}(\varphi )\left| \frac{\partial
R_{OF}(\varphi )}{\partial \varphi }\right| ^{-1}$. The minimum
uncertainty is achieved for $\varphi =\frac{\pi }{2}$. The
resulting sensitivity averaged over $N$ trials is thus
$S_{OF}=V\sqrt{R_{mean}}\,\sqrt{N}$, where $V$ is the visibility of the
fringe pattern. This expression shows that
the phase fluctuations no longer depend on the efficiency $\eta $
of the channel, but only on the average percentage of detected pulses $%
R_{mean}$.\\
\indent We have experimentally tested the enhancement obtained by the OF
strategy. We report in Fig. \ref{fig:experiment}-(b) the
experimental trend of the enhancement as a function of the signal
rate compared with the expected theoretical trend ($p=0.14,
\eta=0.005$). In the adopted apparatus the single-photon fringe
pattern shows a visibility $\sim 50\%$  due to the generation of
more than a single-photon pair by the first nonlinear crystal
adopted as the heralded single-photon source. This seed visibility
value is also responsible for a reduction of the amplified state
visibility and has been taken into account in the comparison
between the two strategies. By comparing the enhancement obtained
through the counting and the OF-based detection methods we can
conclude that the first one allows us to achieve a higher
enhancement.\\
\indent In conclusion, the ability to generate suitable quantum light
probes and quantum detectors is a crucial prerequisite for the
operation of any quantum sensor. The optimal probes maximizing the
sensitivity and performance of the sensors can be theoretically
determined, but the resulting quantum states are often very
complicated, difficult to generate, and extremely sensitive to
losses and noise. We proposed and realized  a simple conceptual strategy to
apply in a lossy scenario that can be engineered with the existing
quantum-optics technology.  Our results show that a large
sensitivity improvement can be achieved even in a high losses
condition in which the dominant losses act after the interaction of
the probe with the sample, hence including all the inefficiencies
in the detection of the probe (spatial and spectral filtering,
transmission, efficiency of the detectors).

We acknowledge support by the ``Futuro in Ricerca'' Project HYTEQ,
and Progetto d'Ateneo of ``Sapienza'' Universit\`{a} di Roma.

%\bibliographystyle{h-physrev}
%\bibliography{bibliography}

\begin{thebibliography}{10}

\bibitem{Giov04b}
V.~Giovannetti {\em et~al.}, 
\newblock Science {\bf 306}, 1330 (2004).

\bibitem{Giov06}
V.~Giovannetti {\em et~al.},
\newblock Phys. Rev. Lett. {\bf 96}, 010401 (2006).

\bibitem{Hels76}
C.~W. Helstrom,
\newblock {\em Quantum Detection and Estimation Theory} (Academic Press, New York, 1976).

\bibitem{Boto00}
A.~N. Boto {\em et~al.},
\newblock Phys. Rev. Lett. {\bf 85}, 2733 (2000).

\bibitem{Dowl08}
J.~P. Dowling,
\newblock Contemp. Phys. {\bf 49}, 125 (2008).

\bibitem{Dang01}
M.~{D'Angelo} {\em et~al.},
\newblock Phys. Rev. Lett. {\bf 87}, 013602 (2001).

\bibitem{Walt04}
P.~Walther {\em et~al.},
\newblock Nature {\bf 429}, 158 (2004).

\bibitem{Mitc04}
M.~W. Mitchell {\em et~al.},
\newblock Nature {\bf 429}, 161 (2004).

\bibitem{Eise05}
H.~S. Eisenberg {\em et~al.},
\newblock Phys. Rev. Lett. {\bf 94}, 090502 (2005).

\bibitem{Naga07a}
T.~Nagata {\em et~al.},
\newblock Science {\bf 316}, 726 (2007).

\bibitem{Ono10}
T. Ono {\em et~al.},
\newblock Phys. Rev. A {\bf 81}, 033819 (2010).

\bibitem{Afek10}
I. Afek {\em et~al.},
\newblock Science {\bf 328}, 879 (2010).

\bibitem{Gilb08}
G. Gilbert {\em et~al.},
\newblock J. Mod. Opt. {\bf 55}, 3283 (2008).

\bibitem{Rubi07}
M.~A. Rubin {\em et~al.},
\newblock Phys. Rev. A {\bf 75}, 053805 (2007).

\bibitem{Shaj07}
A.~Shaji {\em et~al.},
\newblock Phys. Rev. A {\bf 76}, 032111 (2007).

\bibitem{Huve08}
S.~Huver {\em et~al.},
\newblock Phys. Rev. A {\bf 78}, 063828 (2008).

\bibitem{Dorn09}
U.~Dorner {\em et~al.},
\newblock Phys. Rev. Lett. {\bf 102}, 040403 (2009).

\bibitem{Macc09}
L.~Maccone {\em et~al.},
\newblock Phys. Rev. A {\bf 79}, 023812 (2009).

\bibitem{Demk09}
R.~Demkowicz-Dobrzanski {\em et~al.},
\newblock Phys. Rev. A {\bf 80}, 013825 (2009).

\bibitem{Kacp09}
M.~Kacprowicz {\em et~al.},
\newblock Nature Photonics, {\bf 4}, 357 (2010).

\bibitem{Lee09}
T.-W. Lee {\em et~al.},
\newblock Phys. Rev. A {\bf 80}, 063803 (2009).

\bibitem{DeMa98}
F.~{De Martini},
\newblock Phys. Rev. Lett. {\bf 81}, 2842 (1998).

\bibitem{DeMa98a}
F.~{De Martini},
\newblock Phys. Lett. A {\bf 250}, 15 (1998).

\bibitem{DeMa08}
F.~{De Martini} {\em et~al.},
\newblock Phys. Rev. Lett. {\bf 100}, 253601 (2008).

\bibitem{DeMa09}
F.~{De Martini} {\em et~al.},
\newblock Phys. Rev. A {\bf 79}, 052305 (2009).

\bibitem{DeMa09a}
F.~{De Martini} {\em et~al.},
\newblock Phys. Rev. Lett. {\bf 103}, 100501 (2009).

\bibitem{Naga07}
E.~Nagali {\em et~al.},
\newblock Phys. Rev. A {\bf 76}, 042126 (2007).

\bibitem{Leve97}
J.~A.~Levenson {\em et~al.},
\newblock Quantum Semiclass. Opt. {\bf 9}, 221 (1997).

\bibitem{Glau66}
B.~R.~Mollow {\em et~al.},
\newblock Phys. Rev. {\bf 160}, 1076 (1966); Phys. Rev. 
{\bf 160}, 1097 (1966).

\bibitem{supplementary} Supplementary Information.

\bibitem{Pari09}
M.~G.~A. Paris,
\newblock Int. J. Quant. Inf. {\bf 7}, 125 (2009).

\end{thebibliography}

\end{document}

\title{Supplementary informations: Enhanced resolution of lossy interferometry by coherent amplification of single photons}

\author{Chiara Vitelli}
\affiliation{Dipartimento di Fisica, ``Sapienza'' Universit\`{a} di Roma,
piazzale Aldo Moro 5, I-00185 Roma, Italy}
\affiliation{Consorzio Nazionale Interuniversitario per le Scienze Fisiche
della Materia, piazzale Aldo Moro 5, I-00185 Roma, Italy}

\author{Nicol\`{o} Spagnolo}
\affiliation{Dipartimento di Fisica, ``Sapienza'' Universit\`{a} di Roma,
piazzale Aldo Moro 5, I-00185 Roma, Italy}
\affiliation{Consorzio Nazionale Interuniversitario per le Scienze Fisiche
della Materia, piazzale Aldo Moro 5, I-00185 Roma, Italy}

\author{Lorenzo Toffoli}
\affiliation{Dipartimento di Fisica, ``Sapienza'' Universit\`{a} di Roma,
piazzale Aldo Moro 5, I-00185 Roma, Italy}

\author{Fabio Sciarrino}
\email{fabio.sciarrino@uniroma1.it}
\affiliation{Dipartimento di Fisica, ``Sapienza'' Universit\`{a} di Roma,
piazzale Aldo Moro 5, I-00185 Roma, Italy}
\affiliation{Istituto Nazionale
di Ottica Applicata, largo Fermi 6, I-50125 Firenze, Italy}

\author{Francesco De Martini}
\affiliation{Dipartimento di Fisica, ``Sapienza'' Universit\`{a} di Roma,
piazzale Aldo Moro 5, I-00185 Roma, Italy}
\affiliation{Accademia Nazionale dei Lincei, via della Lungara 10, I-00165 Roma,
Italy}

\maketitle
\section{Experimental implementation of the protocol}
\label{sec:exp_SPCM}
\begin{widetext}

\begin{figure}[ht]
\centering
\includegraphics[width=0.8\textwidth]{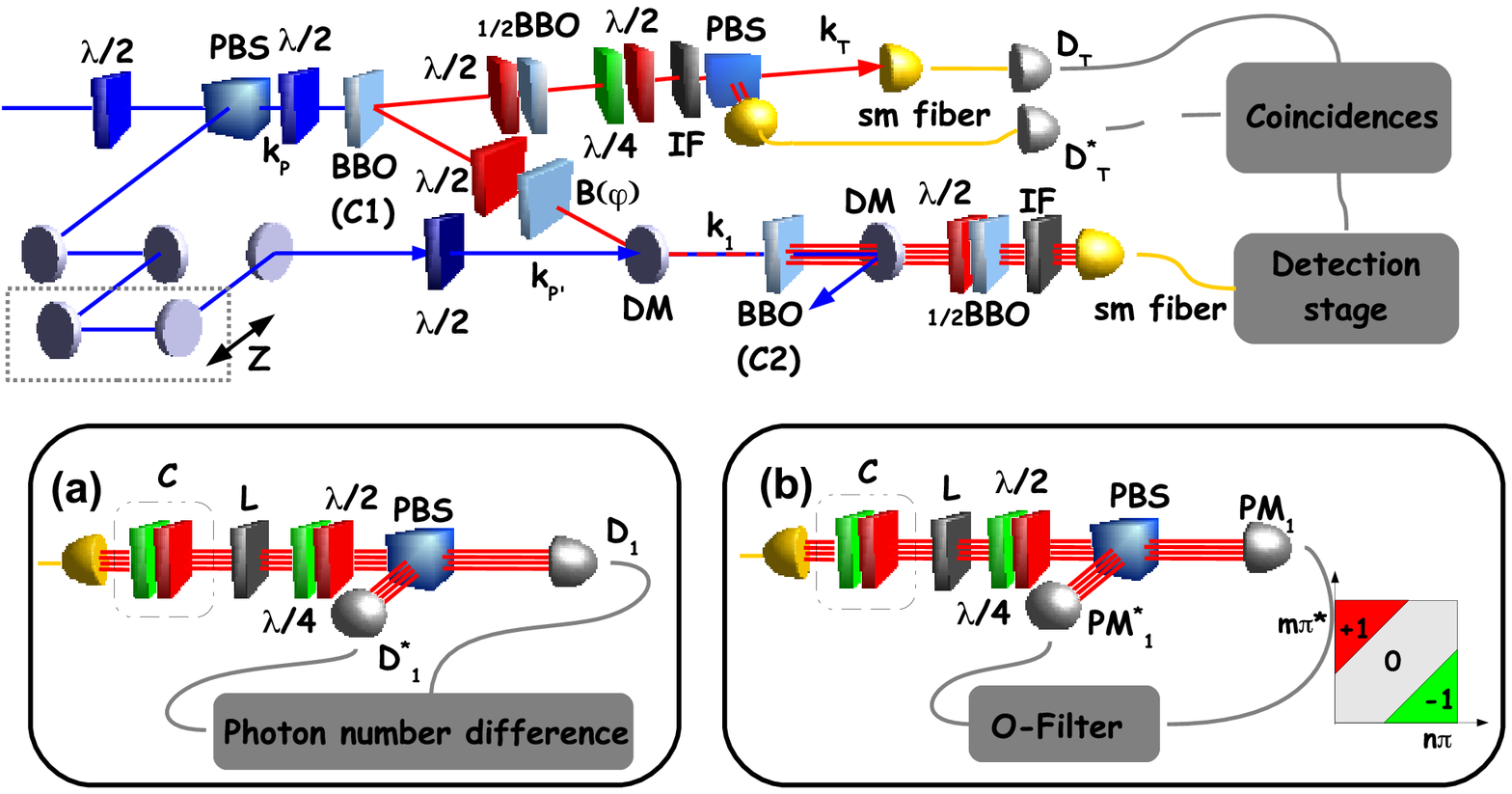}
\caption{Experimental setup of the QI-OPA based interferometric scheme. An intense pulsed
pump beam of $P=750$ mW and $\lambda = 397.5$ nm is generated by a Ti:Sa laser system.
The pump beam is split between modes $\mathbf{k}_{p}$ and $\mathbf{k}'_{p}$.
In the first BBO crystal (C1), a single photon $\vert + \rangle_{1}$ state is prepared upon
detection on mode $\mathbf{k}_{T}$ with a single-photon SPCM-AQR14 detector $(D_{T})$.
Then, the phase $\varphi$ is introduced through a Babinet-Soleil compensator (B) and
the probe state $\vert \varphi \rangle$ after the interaction is then injected in the QI-OPA
and superimposed with the pump beam on mode $\mathbf{k}'_{p}$. Spatial and temporal
matching are obtained through an adjustable delay line (Z). After the amplification process has occured,
spectral and spatial filtering are performed through the insertion of an interference filter (IF) and
the coupling with a single-mode fiber. The signal is then sent to the detection apparatus. (a) Measurement
scheme based on detection of the average of the photon number difference $\hat{D} =
\hat{n}_{+} - \hat{n}_{-}$. After fiber polarization compensation (C), the output field undergoes
controlled losses up to the single-photon level throughout a tunable attenuator (L). Then, the field is analyzed in
polarization and detected with two single-photon SPCM-AQR14 detectors $(D_{1},D_{1}^{\ast})$ (b)
O-Filter based detection scheme. After fiber polarization compensation (C), the field is analyzed in
polarization and detected with two photomultipiers $(PM_{1},PM_{1}^{\ast})$. The
photocurrents are shot-by-shot processed by an electronic device (O-Filter), whose action
is described in the text and in Ref.\cite{DeMa08}.}
\label{fig:experimental_setup}
\end{figure}

\end{widetext}

We address in more details the experimental implementation of the proposed metrology scheme: 
figure \ref{fig:experimental_setup}. The single photon probe is conditionally 
generated by a BBO non linear crystal (C1) on spatial mode $\mathbf{k}_{1}$ together with 
an entangled photon on spatial mode $\mathbf{k}_{T}$. The overall state produced by $C1$ 
is expressed by: $\vert \psi^{-} \rangle = \frac{1}{\sqrt{2}} \left( \vert H \rangle_{T} 
\vert V \rangle_{1} - \vert V \rangle_{T} \vert H \rangle_{1}\right)$, and is obtained 
through the process of spontaneous parametric down-conversion, consisting in the annihilation
of a pump photon at wavelength (wl) $\lambda_{p}=397.5$ nm followed by the creation of twin 
photons, orthogonally polarized, at wl $\lambda=795$ nm. 

\begin{figure}[ht]
\includegraphics[scale=.8]{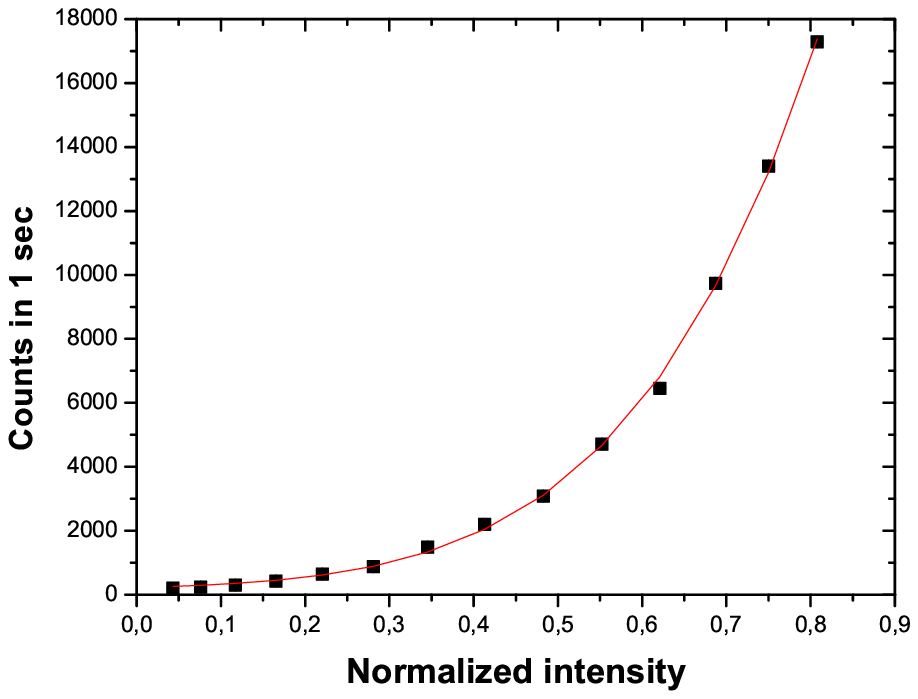}
\caption{Trend of counts detected by $D_{1},D_{1}^{*}$ as a function of the normalized 
intensity signal impinging on the amplifier, working in a spontaneous (not-injected) 
configuration.}
\label{fig:gain}
\end{figure}

The overall pump beam is generated by a Ti:Sa laser system. At a first stage, a pulsed 250 fs 
beam with a repetion rate of 75 MHz at wavelength $\lambda = 795$ nm is generated Ti:Sa modelocked 
MIRA900, pumped by a Verdi V5 Nd:Yag solid state laser. The output beam from the MIRA900 is 
injected into the Ti:Sa REGA9000 amplifier, pumped by a Verdi V18 with an optical power set to 14 W. The overall laser system 
allows to obtain a $1.5$ W output beam at wavelength $\lambda=795$ nm, that, after a double 
harmonic generation process, generates the experiment pump beam at wl $\lambda_{p}=397.5$ nm 
of power $P=750$ W. This beam is split into two parts with different optical powers; the weaker 
one impinges into $C1$ and generates the entangled state $\vert \psi^{-} \rangle$, the stronger 
one is directed towards the amplifier crystal $C2$ and is superimposed spatially and temporally 
with the single photon probe after the interaction with the sample (simulated by a Soleil-Babinet compansator
$B(\varphi)$, which allows to insert a tunable phase shift between the $\vec{\pi}_{H}$ and the $\vec{\pi}_{V}$ 
polarization, in figure \ref{fig:experimental_setup}). The number of photons generated in the 
amplification process depends exponentially on the nonlinear gain of the amplifier $g$. The 
maximum value of $g$ experimentally found is $g_{max}=4.5$. The value of the nonlinear gain 
has been measured by exploiting the method of Ref.\cite{Eise04,Cami06}, i.e. by measuring the dependence 
of the counts rate on the incident pump power in the spontaneous regime and by fitting the curve 
with the expected functional form $C = \frac{\eta \tanh^2 g}{1-(1-\eta)\tanh^2 g}$ [Fig.\ref{fig:gain}].

\subsection{High losses regime and single-photon detection}\label{sec:SPCM}
In this section we investigate the first measurement strategy implemented in our experiment 
[Fig.\ref{fig:experimental_setup}-(a)]. The amplified state, after spectral and spatial 
filtering, is deliberately attenuated up to the single-photon level. The signal is then 
analyzed in polarization and detected by two single photon counting modules SPCM, $D_{1}$ 
and $D_{1}^{*}$ in figure \ref{fig:experimental_setup}-(a). The resulting signals, triggered 
by the click of detectors $\{ D_{T},D_{T}^{*} \}$ on mode $\mathbf{k}_{T}$, are hence 
subtracted and the difference in the number of photons $\hat{D}$ is recorded as a function 
of the phase $\varphi$, inserted by the babinet $B(\varphi)$ on the probe path.

\begin{figure}[ht]
\includegraphics[width=0.35\textwidth]{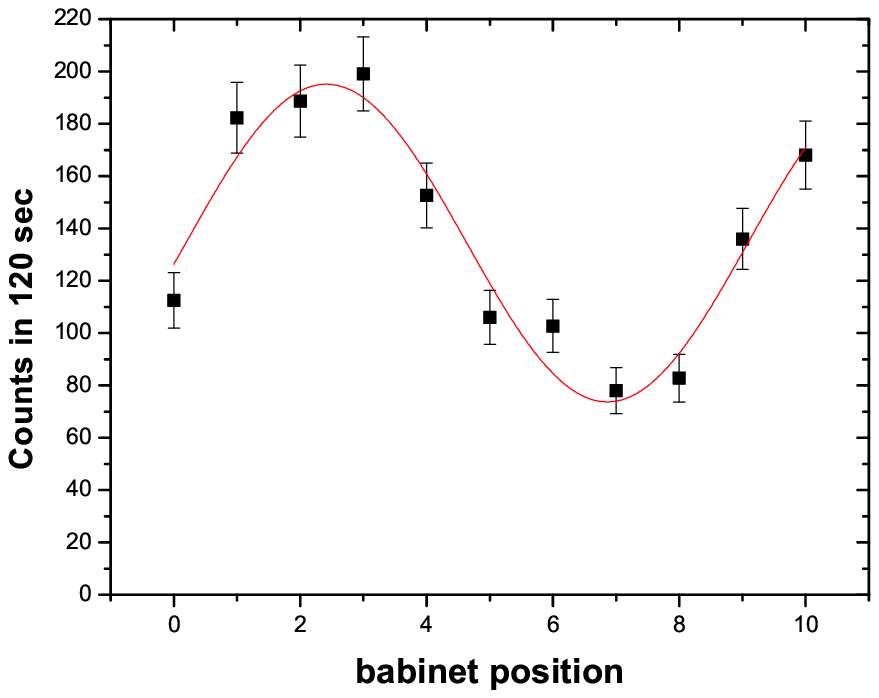}
\includegraphics[width=0.35\textwidth]{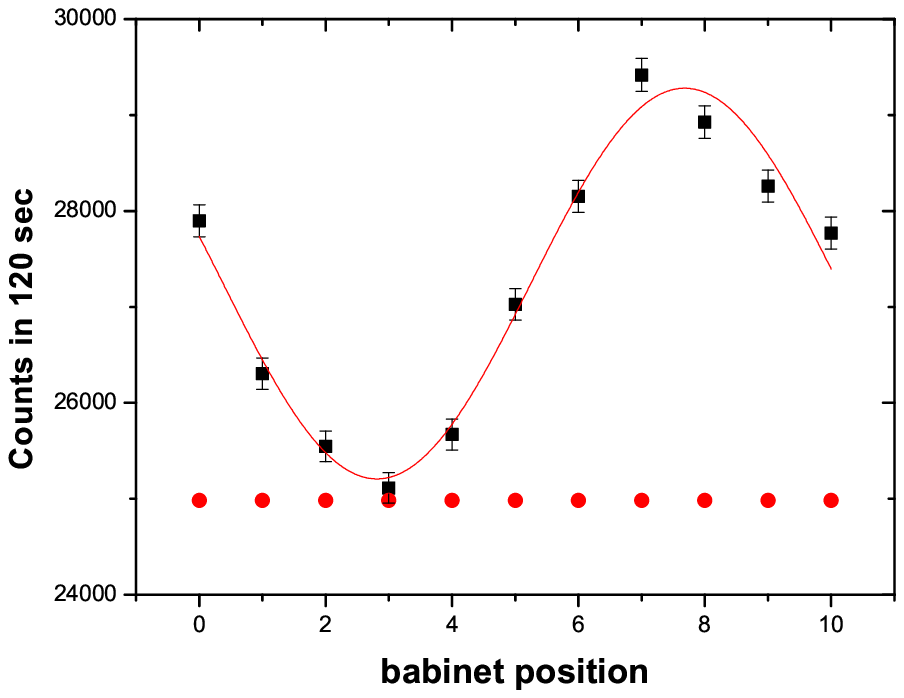}
\caption{Experimental fringe pattern for the single photon probe (left) and for the amplified beam (right).}
\label{fig:fringe_comparison}
\end{figure}

\noindent The fringe patterns obtained in absence and in presence of the amplifier are 
reported in Fig.\ref{fig:fringe_comparison}, in order to compare the performances of the 
two phase-estimation schemes. In the first case the obtained visibility is $V_{1phot}=0.45$, 
whose value differs from the expected unitary one due to the generation of more than a single 
photon pair by the first nonlinear crystal, adopted as the heralded single photon source. This 
value of the seed visibility is also responsible for a reduction of the amplified state visibility 
$V_{ampl}(g_{max})=0.08$ and has been taken into account in the comparison between the two 
strategies. 

\subsection{Increased resilience to losses through an orthogonality filter}\label{sec:OF}
In this section we investigate an alternative measurement strategy apt to
obtain an enhancement in the resilience to losses of the interferometric
estimation of an unknown phase. The presented method is based on an electronic 
discriminator, the orthogonality filter(OF) device introduced in Ref.\cite{DeMa08}, 
which performs a dichotomic threshold detection, allowing to increase the visibility 
of the amplified signal with respect to the SPCM-based strategy at the cost
of a lower detection rate.

The action of the OF is then to select those events which can be discriminated
with higher fidelity, leading to an increase in the visibility. The latter operation
is achieved at the cost of discarding a part of the data. According to these
considerations, the quantum efficiency of the scheme is only given by the average signal
$\overline{\eta} = R_{mean}(k)$ filtered by the OF device.
This parameter $\overline{\eta}$ takes the role of the overall efficiency of the
OF-based scheme.
The sensitivity, found through the standard
definition, is: $\frac{1}{\Delta \varphi} = \frac{1}{\Delta I_{OF}(\varphi)} \left \vert
\frac{\partial I_{OF}(\varphi)}{\partial \varphi} \right \vert$. That can be further symplified
obtaining the expression:
\begin{equation}
S_{OF}=\frac{1}{\Delta \varphi_{OF}} = \frac{\left \vert \sin \varphi \, \left( I_{\mathrm{max}} 
- I_{\mathrm{min}} \right)
\right \vert}{\left( (I_{\mathrm{max}} - I_{\mathrm{min}}) \cos^{2}
\frac{\varphi}{2} + I_{\mathrm{min}} \right)^{1/2}}
\end{equation}
The minimum is obtained in $\varphi = \frac{\pi}{2}$, where the sensitivity can be
put in the form:
\begin{equation}
S_{OF}=\frac{1}{\Delta \varphi_{OF}} = V \sqrt{R_{mean}}
\end{equation}
Finally, the average over $N$ trials gives an improvement of $\sqrt{N}$:
\begin{equation}
S_{OF} = V \sqrt{R_{mean}}\,\sqrt{N}
\end{equation}
This expression shows that the sensitivity $S_{OF}$ does not depend on
the efficiency $\eta$ of the channel, but only on the average filtered signal $R_{mean}$.

\begin{figure}[ht]
\includegraphics[width=0.4\textwidth]{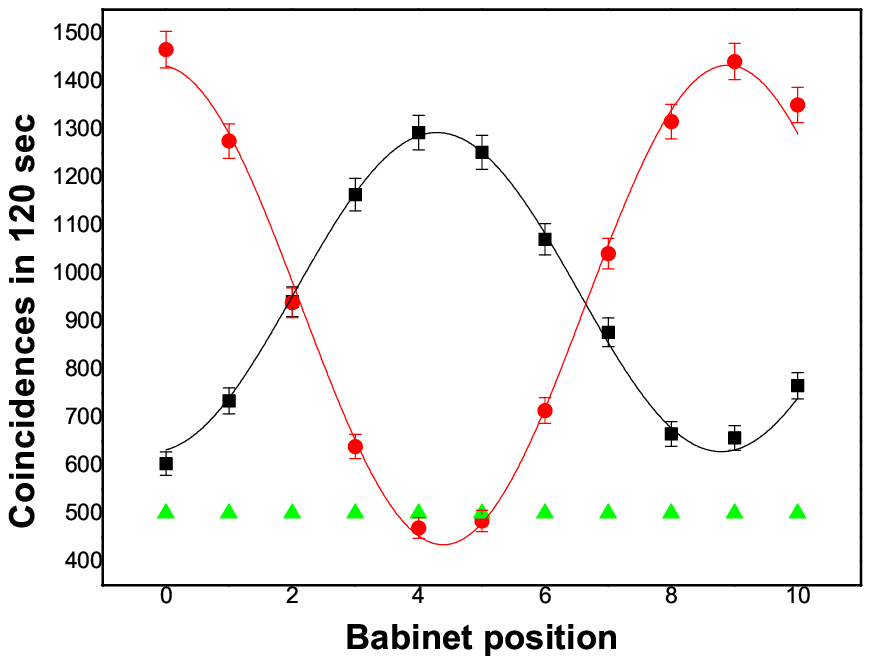}
\caption{Experimental fringe pattern for a percentage of detected signal equal to $3.6\times 10^{4}$, 
the obtained visibility is $\sim 0.53$.} \label{fig:experiment_OF_fringe}
\end{figure}

The OF based strategy has been experimentally implemented by adopting at the detection stage two 
photomultipliers, $PM_{1}$ and $PM_{1}^{*}$ in figure \ref{fig:experimental_setup}-(b).  The two 
intensity signals, proportional to the orthogonally polarized number of photons, are compared 
shot-by-shot by the orthogonality-filter that renders at its output two TTL signals. Coincidences 
of such signals with the corresponding click of the detector on trigger mode ($D_{T}$ or $D_{T}^{*}$) 
are recorded. By increasing the discrimination threshold of the OF we can achieve a higher visibility 
for the amplified field respect to the one obtained through the SPCM based strategy of previous 
section. In figure \ref{fig:experiment_OF_fringe} is reported the fringe pattern obtained with a 
percentage of detected signal equal to $3.6\times 10^{-4}$, the resulting visibility is $\sim 0.53$.
\begin{figure}[ht]
\includegraphics[width=0.4\textwidth]{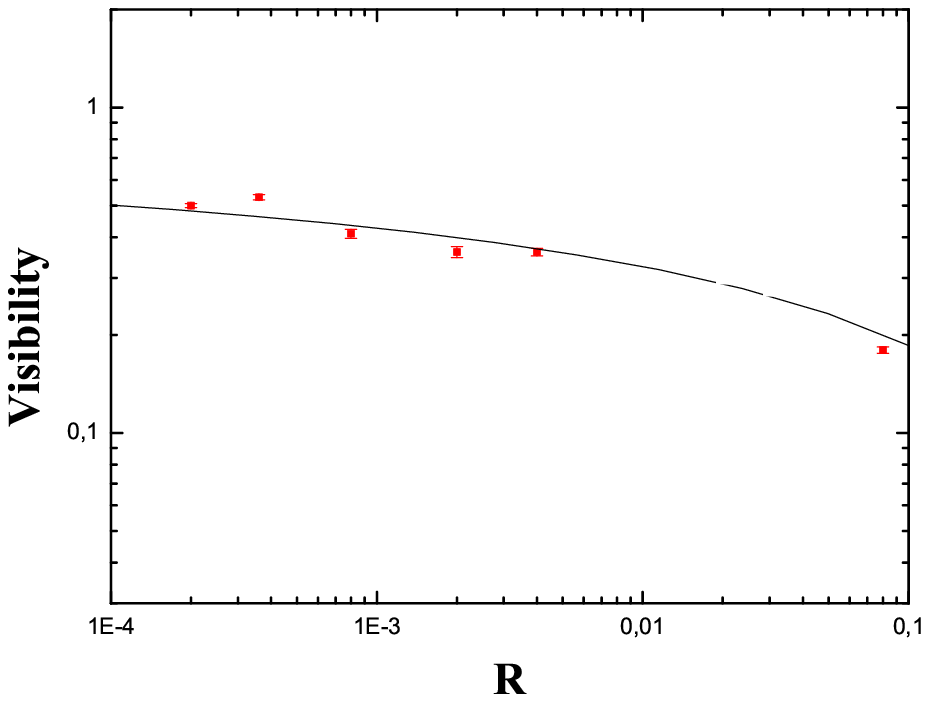}
\caption{Experimental trend of visibility as a function of the percentage of filtered signal, the 
continuous line reports  the theoretical prediction for $p=0.14$, $\eta=0.005$.} 
\label{fig:experiment_OF_visibility}
\end{figure}
The trend of the visibility as a function of the percentage of detected pulses is reported in 
figure \ref{fig:experiment_OF_visibility}, compared with the theoretical prediction for $p=0.14$ 
and $\eta=0.005$. We observe that a higher visibility is obtained for a lower count rate.
By comparing the fringe pattern in figure \ref{fig:experiment_OF_fringe} with the one in figure 
\ref{fig:fringe_comparison}-(b), we observe that the amplified field visibility is increased by 
a factor $\sim 7$ but the counts rate is decreased by a factor $30$. It is worth noting that the 
two detection strategies implemented in this paper (single-photon vs OF) have been investigated
in two different regimes. For the SPCM-based strategy, a tunable attenuator has been exploited to 
achieve the condition $\eta \overline{n} \sim 0.1$, corresponding in our experiment to a value
of the overall detection efficiency of $\eta \sim 3 \times 10^{4}$. For the OF detection method, 
the overall efficiency $\eta$ is due to spectral filtering, single-mode fiber coupling, transmission 
losses and detection efficiency of the photomultipliers, leading to an overall value of $\eta \sim 
0.005$.

%\bibliographystyle{h-physrev}
%\bibliography{bibliography}